%% file: main.tex
\documentclass[aps,letterpaper,tightenlines,notitlepage,superscriptaddress,12pt]{revtex4-1}
\usepackage{fullpage,graphicx,psfrag,verbatim}
\usepackage{amsmath}
\usepackage{amsfonts}
\usepackage{amssymb}
\usepackage{bm}
\usepackage[colorlinks]{hyperref}
\usepackage{xcolor}
\usepackage{subfig}

\input{defs.tex}

\begin{document}
\title{Photon Energy-Resolved Velocity Map Imaging from Spectral Domain Ghost Imaging}
\author{Jun Wang}
\email{junwang9@stanford.edu}
\affiliation{Stanford PULSE Institute, SLAC National Accelerator Laboratory, Menlo Park, CA 94025, USA}
\affiliation{Department of Applied Physics, Stanford University, Stanford, CA 94305, USA}
\author{Taran Driver}
\email{tdriver@stanford.edu}
\affiliation{Stanford PULSE Institute, SLAC National Accelerator Laboratory, Menlo Park, CA 94025, USA}
\affiliation{SLAC National Accelerator Laboratory, Menlo Park, CA 94025, USA}

\author{Felix Allum}
\affiliation{Stanford PULSE Institute, SLAC National Accelerator Laboratory, Menlo Park, CA 94025, USA}
\author{Christina C. Papadopoulou}
\affiliation{Deutsches Elektronen-Synchrotron DESY, Notkestr. 85, 22607 Hamburg, Germany}
\author{Christopher Passow}
\affiliation{Deutsches Elektronen-Synchrotron DESY, Notkestr. 85, 22607 Hamburg, Germany}
\author{Günter Brenner}
\affiliation{Deutsches Elektronen-Synchrotron DESY, Notkestr. 85, 22607 Hamburg, Germany}
\author{Siqi Li}
\affiliation{SLAC National Accelerator Laboratory, Menlo Park, CA 94025, USA}
\author{Stefan Düsterer}
\affiliation{Deutsches Elektronen-Synchrotron DESY, Notkestr. 85, 22607 Hamburg, Germany}
\author{Atia Tul Noor}
\affiliation{Deutsches Elektronen-Synchrotron DESY, Notkestr. 85, 22607 Hamburg, Germany}
\author{Sonu Kumar}
\affiliation{Deutsches Elektronen-Synchrotron DESY, Notkestr. 85, 22607 Hamburg, Germany}
\author{Philip H. Bucksbaum}
\affiliation{Stanford PULSE Institute, SLAC National Accelerator Laboratory, Menlo Park, CA 94025, USA}
\affiliation{SLAC National Accelerator Laboratory, Menlo Park, CA 94025, USA}
\author{Benjamin Erk}
\affiliation{Deutsches Elektronen-Synchrotron DESY, Notkestr. 85, 22607 Hamburg, Germany}
\author{Ruaridh Forbes}
\affiliation{Stanford PULSE Institute, SLAC National Accelerator Laboratory, Menlo Park, CA 94025, USA}
\affiliation{SLAC National Accelerator Laboratory, Menlo Park, CA 94025, USA}
\author{James P. Cryan}
\email{jcryan@slac.stanford.edu}
\affiliation{Stanford PULSE Institute, SLAC National Accelerator Laboratory, Menlo Park, CA 94025, USA}
\affiliation{SLAC National Accelerator Laboratory, Menlo Park, CA 94025, USA}

\date{Oct 2022}

\begin{abstract}
    We present an approach that combines photon spectrum correlation analysis with the reconstruction of three-dimensional momentum distribution from velocity map images in an efficient, single-step procedure.
    We demonstrate its efficacy with the results from the photoionization of the $2p$-shell of argon using the FLASH free-electron laser~(FEL). 
    Distinct spectral features due to the spin-orbit splitting of Ar$^+(2p^{-1})$ are resolved, despite the large average bandwidth of the ionizing pulses from the FEL.
    This demonstrates a clear advantage over the conventional analysis method, and it will be broadly beneficial for velocity map imaging experiments with FEL sources.
    The retrieved linewidth of the binding energy spectrum approaches the resolution limitation prescribed by the spectrometers used to collect the data.
    Our approach presents a path to extend spectral-domain ghost imaging to the case where the photoproduct observable is high-dimensional.
\end{abstract}
\maketitle

\section{Introduction}

Modern x-ray spectroscopy provides a sensitive probe of local electronic density in molecular systems with atomic-site specificity~\cite{siegbahn_electron_1982, picon_hetero-site-specific_2016, erk_imaging_2014}. %
As a result, x-ray free-electron lasers~(XFELs) with continuous wavelength tunability throughout the soft x-ray region, unparalleled peak brightness, and short temporal duration, have enabled significant advances in time-resolved measurements of molecular dynamics~\cite{wolf_probing_2017, jay_capturing_2022, brause_time-resolved_2018, allum_loc_2022}.
Oftentimes the intrinsic photon energy jitter of a self-amplified spontaneous emission~(SASE) XFEL, along with the strongly fluctuating sub-structure, is thought to be the limiting factor for achievable energy resolution in time-resolved spectroscopy~\cite{brause_time-resolved_2018, allum_loc_2022, mayer2022following}.
However, the spectral fluctuations inherent to SASE operation can, in fact, be exploited as a notable advantage by correlating x-ray observables with properties of the incident pulse on a shot-to-shot basis.
This is a powerful approach to improving the resolution of experiments at XFEL facilities, beyond the bandwidth limit.
In particular, the application of the spectral-domain ghost imaging~(SDGI) technique has demonstrated sub-bandwidth resolution for spectroscopies employing  XFELs~\cite{DriverPCCP,Li_JPB2021,kayser2019core, Klein_HighResolution_2022}.

Thanks to their high throughput, $4\pi$ collection solid-angle, and ability to provide angle-resolved photoproduct yields, velocity map imaging~(VMI) spectrometers~\cite{Eppink1997VMI} have emerged as a popular instrument for time-resolved XFEL studies of dynamics in gas phase systems~\cite{erk_imaging_2014, squibb_acetylacetone_2018, Eppink1997VMI, Rouzee_Angle-resolved2011, Mizuho_TRPE_JCP_2021, prince_coherent_2016, brause_time-resolved_2018,allum_loc_2022}.
A VMI spectrometer typically measures the two-dimensional~(2D) projection of the three-dimensional~(3D) momentum distribution of charged particles.
In most cases, however, the quantity of interest is the underlying 3D distribution or another related quantity
such as the kinetic energy~(KE) spectrum, access to which requires post-processing of the 2D projection that inverts the Abel transform \cite{Gustavo2004pBASEX, Roberts2009OnionPeeling, Dick2014MEVIR, MONTGOMERYSMITH1988AbelInversion, Vrakking2001IterAbelInv}.
In trying to apply SDGI to retrieve the photon energy-resolved KE spectrum of charged particles from a VMI data set, the momentum projection, inherent to the VMI operation principal, could complicate the procedure.  
A feature with a specific KE is mapped to a broad range of pixels on the VMI detector, and thus features with distinct KEs will produce heavily overlapping features in the projected distribution
This is in contrast with previous applications of SDGI, where KE is more directly related to the experimental measurement~\cite{Li_JPB2021}. 
In this work, we present a single-step regression approach which can simultaneously reconstruct the spectral response and the 3D momentum distribution by exploiting correlation between the single-shot VMI image and the corresponding photon spectrum demonstrating that the projection inherent in the VMI concept is not detrimental to the SDGI procedure. 

\begin{figure}[hbtp]
    \centering
    \includegraphics[width=0.7\linewidth]{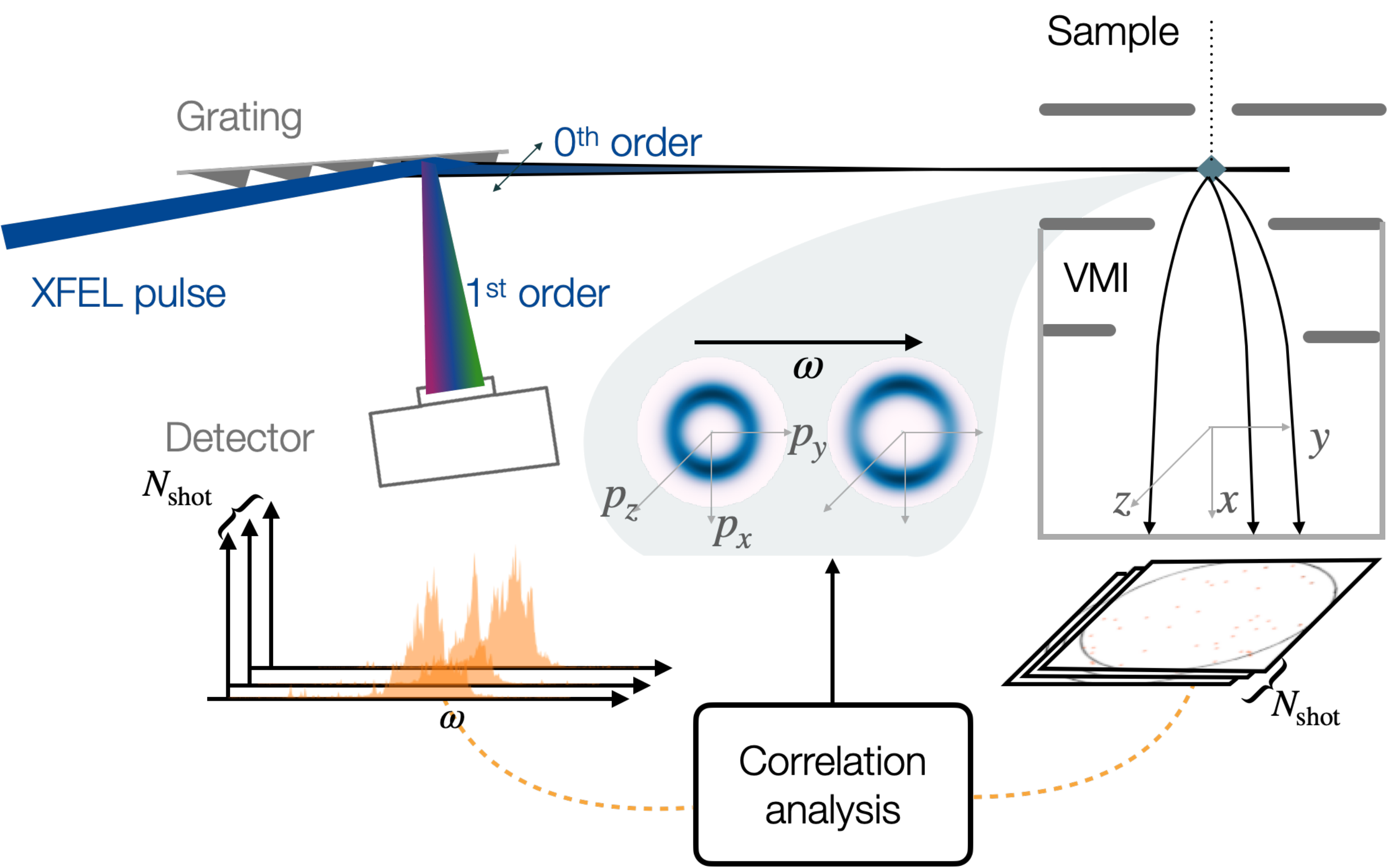}
    \caption{Schematic of a typical experimental layout that correlation analysis approaches can be applied to. 
    XFEL pulses are incident on gaseous sample. 
    The vital observables for our approach are
    the charged particles' VMI images and the x-ray spectra simultaneously measured with a high-resolution photon spectrometer on a shot-to-shot basis. Throughout this work, we refer the VMI axis as $x$ direction, and in this figure the x-ray is polarized along $z$ direction and propagates along $y$.
    }
    \label{fig:measurement}
\end{figure}

The primary components of a typical experiment where correlation analysis can be applied is illustrated in Fig.~\ref{fig:measurement}. 
Briefly, a stochastic light source~(here the SASE XFEL~\cite{feldhaus2010flash}) produces pulses that intercept a grating. 
The reflected beam~(zeroth order) is directed toward the interaction point of a VMI spectrometer, which records the single-shot photoelectron momentum distribution.
The incident spectrum is measured from the first order diffraction of the grating~\cite{brenner2011first}. %

In Sec.~\ref{sec:method}, we describe our single-step regression approach. 
In Sec.~\ref{sec:results}, we demonstrate it with an experiment at the Free-electron LASer in Hamburg (FLASH), where spectral resolution below the average bandwidth of the incident x-ray pulses has been achieved.
We point out that sub-bandwidth resolution is a common advantage of photon spectrum correlation analysis approaches.
In Sec.~\ref{sec:Discussion}, we describe the advantage we observe in regularizing with a single-step reconstruction compared to other possible implementations. 
Our approach promises to enhance the energy resolution of XFEL experiments using VMI, opening up opportunities to study time-dependent phenomena through the variation in finely resolved energy structures \cite{inhester_spectroscopic_2019, allum_loc_2022}.

\section{Method}\label{sec:method}
\subsection{Model}
The primary quantity of interest in VMI measurements %
is the 3D momentum distribution of charged particles, $f(\bm{p})$, 
which depends on the spectral profile of the incoming pulse,
according to 
\begin{equation}
f(\bm{p}) = \int \chi(\bm{p},\omega)a(\omega)\d\omega,
\end{equation}
where $a(\omega)$ is the frequency spectrum of an incident pulse, and $\chi(\bm{p},\omega)$ is the~(linear) response to the spectral intensity at frequency $\omega$.
This model applies to light-matter interactions in the linear regime, such as single-photon ionization. %
The image obtained in a VMI spectrometer is given by the Abel transform of $f$ with a mapping from momentum to position $\bm{r}=\bm{p}/\sqrt{2m\alpha}$,
\begin{equation} \label{eq:model-continuous}
    b(y,z) = \int \tilde{\chi}(\bm{r},\omega)a(\omega)\d x\d\omega~,
\end{equation}
where $x$ is the direction of projection, $\alpha\propto mt_\mathrm{ToF}^{-2}$ is the magnification factor depending on the time-of-flight $t_\mathrm{ToF}$ and particle mass $m$, and $\tilde{\chi}(\bm{r},\omega) = (2m\alpha)^{3/2} \chi(\bm{p},\omega)$ is the $\omega$-dependent momentum distribution.%

\begin{figure}[hbtp]
    \centering
    \includegraphics[width=0.7\linewidth]{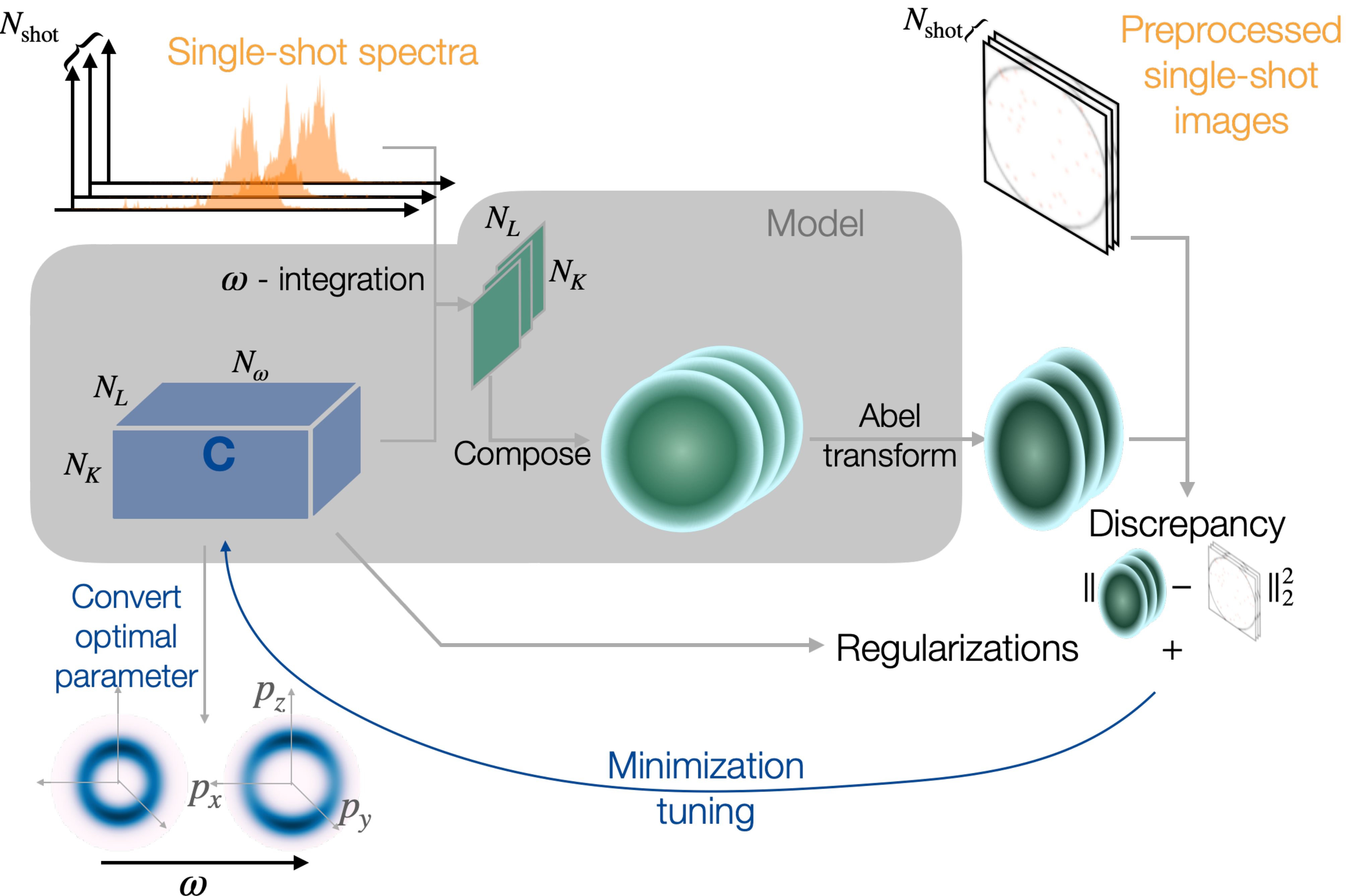}
    \caption{Schematic of the overall analysis procedure. Given a single-shot spectral profile, our model composes a 3D momentum distribution based on parameters $C$ and conducts the Abel transform, obtaining an expected 2D distribution to be compared with the corresponding VMI image. 
    Tuning the model parameters arrives at an optimal point of the objective function that consists the total discrepancy and the regularization terms. 
    The optimal set of parameters is eventually converted to the photon-frequency dependent momentum distribution.}
    \label{fig:method-schematic}
\end{figure}

Our analysis approach is summarized graphically in Fig.~\ref{fig:method-schematic}. 
The spectral response of the sample, $\chi(\bm{p},\omega)$, is extracted by fitting a model to a combination of single-shot VMI images and photon spectra. 
The most likely $\tilde{\chi}$ is the function that minimizes the difference between the measured and predicted electron momentum distributions, subject to regularization terms that favor sparsity and smoothness in the $\omega$-dependent distribution $\tilde{\chi}$. 
We restrict our considerations to cases with cylindrical symmetry, where the axis of symmetry~($z$), is oriented perpendicular to the direction of projection~($x$), such as the case shown in Fig.~\ref{fig:measurement}. 
Other configurations satisfying the same symmetry requirements are discussed in Sec. \ref{sec:configs} .
Such a model is sufficient to describe experiments using either linearly- or circularly-polarized laser pulses acting on isotropic samples. 
The Abel transform is uniquely invertible under this symmetry condition, and the quality of the reconstruction will depend on the number of FEL shots used.

\subsection{Implementation}\label{sec:computational}

In general, the inverse Abel transform is ill-conditioned, being susceptible to experimental noise~\cite{Pretzier1992AbelInversion}. 
Several inversion algorithms~\cite{Gustavo2004pBASEX,Roberts2009OnionPeeling,Dick2014MEVIR, MONTGOMERYSMITH1988AbelInversion, Vrakking2001IterAbelInv} have been developed to mitigate this issue and to robustly reconstruct the underlying 3D momentum distribution. %
In our method we adopt the basis functions employed in the well-known pBasex~\cite{Gustavo2004pBASEX} algorithm to represent $\tilde{\chi}(\bm{r},\omega)$ and combine the inversion procedure with spectral domain ghost imaging~\cite{DriverPCCP,Li_JPB2021}.
pBasex expands the three-dimensional momentum distribution with a basis set that is the product of radial basis functions and Legendre polynomials:
\begin{equation} \label{eq:expand-on-basis}
    \tilde{\chi}(\bm{r},\omega) = \frac{1}{2\pi}\sum_{l=0}^{N_L}\sum_{k=1}^{N_K}c_{lk}(\omega)f_k(r)P_{l}(\frac{z}{r})~,
\end{equation}
where $P_{l}(z/r)$ is the $l$-th order Legendre polynomial, %
and the sum over $l$ can be truncated at twice the highest order of light-matter interaction $N_L$ (i.e. twice the number of photons involved in the ionization process).
$\{f_k(r)\}_{k=1}^{N_K}$ is a set of radial basis functions%
, and $c_{lk}(\omega)$ are their $\omega$-dependent coefficients.

Combining Eqn.~(\ref{eq:expand-on-basis}) with Eqn.~(\ref{eq:model-continuous}), the projected 2D distribution can be written as:
\begin{equation}\label{eqn:byz}
    b(y,z) = \sum_{l=0}^{N_L}\sum_{k=1}^{N_K} \int c_{lk}(\omega)G_{lk}(y,z)a(\omega)\d\omega~,
\end{equation}
where $G_{lk}(y,z) = \int f_k(r)P_{l}(z/r)\d x/(2\pi)$, is the projection of each basis function perpendicular to the axis of cylindrical symmetry.
It is worth noting, that after discretization on a Cartesian grid $\bm{R}_q = (y_q,z_q)$, the projections $G_{lk,q}=G_{lk}(y_q,z_q)$ are the same integrals encountered in the standard pBasex inversion method (Eqn.~4 of Ref.~\cite{Gustavo2004pBASEX}). Being further uniformly discretized in photon energy $\omega_w$, 
Eqn.~(\ref{eqn:byz}) is
\begin{equation}
    \hat{B}_{i,q} = \sum_{l=0}^{N_L}\sum_{k=1}^{N_K}\sum_{w=1}^{N_\omega} G_{lk,q}A_{i,w} C_{lk,w}~.
    \label{eqn:B_iq}
\end{equation}
where for each single-shot, $i$, $\hat{B}_{i,q}$ is the expected intensity at pixel $q$ given the spectral profile $A_{i,w}=A_i(\omega_w)$ and the spectral response of the sample encoded in the coefficients $C_{lk,w}=C_{lk}(\omega_w)$. 
Solving for the coefficients $C$ from the expected image $\hat{B}$ is straightforward, as the tensor $H\equiv G\otimes A$ can be reshaped into a matrix $H_{iq,lkw} = G_{lk,q}A_{i,w}$, whose pseudo-inverse maps $\hat{B}$ to $C$. 
However, solving $C$ from the measured images $B$ by application of the pseudo-inverse, is highly sensitive to experimental noise in both $B$ and photon spectra $A$.
A more robust approach is to minimize an objective function, $h$ consisting of $h_0$, which quantifies the model-measurement discrepancy, and regularization terms favoring expected qualities of $C_{lk,w}$, such as sparsity and smoothness~\cite{Li_JPB2021}, which is similar to the procedure discussed in Ref.~\cite{dick_inverting_2013}.
In our case, $h(C)$ is
\begin{equation}\label{eqn:obj}
    h(C) = h_0(C;A,G,B) + \lambda_\mathrm{sp}h_\mathrm{sp}(C) + \lambda_\mathrm{sm,\omega}h_\mathrm{sm,\omega}(C) + \lambda_\mathrm{sm,r}h_\mathrm{sm,r}(C)~,
\end{equation}
where
\begin{equation}
     h_0(C;A,G,B) = \sum_{i} \sum_{q}W_q\left|\hat{B}_{i,q}(C;A,G)- B_{i,q}\right|^2
\end{equation}
is the Gaussian log-likelihood up to a global factor, with weights $W_q$ over each pixel $q$ in the VMI image. 
Weights $W_q$ can be chosen to enhance the sensitivity of the reconstruction to certain regions of the VMI image, and henceforth we denote $W$ to be the diagonal matrix constructed by them.
$h_\mathrm{sp}(C)$ and $h_\mathrm{sm,\cdot}(C)$ are the sparsity and smoothness regularization terms, respectively. 

We separate smoothness into two terms $h_\mathrm{sm,\omega}(C)$ and $h_\mathrm{sm,r}(C)$, to differentiate the smoothness along the frequency~($\omega$), and radial~($r$) directions. 
The corresponding hyperparameters, $\lambda_\mathrm{sp}, \lambda_\mathrm{sm,\omega}, \lambda_\mathrm{sm,r}$ control the degree to which sparsity and smoothness are enforced in the retrieved $\tilde{\chi}$.
For each direction in $d=\omega, r$, the smoothness term quantifies the roughness with the second order difference, i.e. $h_\mathrm{sm,d}(C) =\|L^{(d)}C\|_2^2,~d=\omega,r$, with $L^{(d)}$ representing the finite-difference Laplacian operator along direction $d$.
Common choices for the form of $h_{sp}$ include the $\mathrm{L}_1$-norm \cite{Tibshirani1996Lasso} and $\mathrm{L}_2$-norm squared~\cite{Tikhonov1995}. 
We choose the latter for the demonstration in Sec.~\ref{sec:results}. 
We discuss strategies for choosing the proper values for the hyperparameters in the supplementary material.

From the point of view of implementation, the discrepancy term, $h_0(C; A,G,B)$, is quadratic in the $C$ coefficients, conducting the summation over shot $i$ to formulate $h_0$ into a quadratic form prior to the optimization significantly improves the efficiency, although at the cost of caching a large matrix $A^TA\otimes GWG^T$ in memory.

\section{Results} \label{sec:results}

\begin{figure}[htbp]
    \centering
    \includegraphics[width=0.8\linewidth]{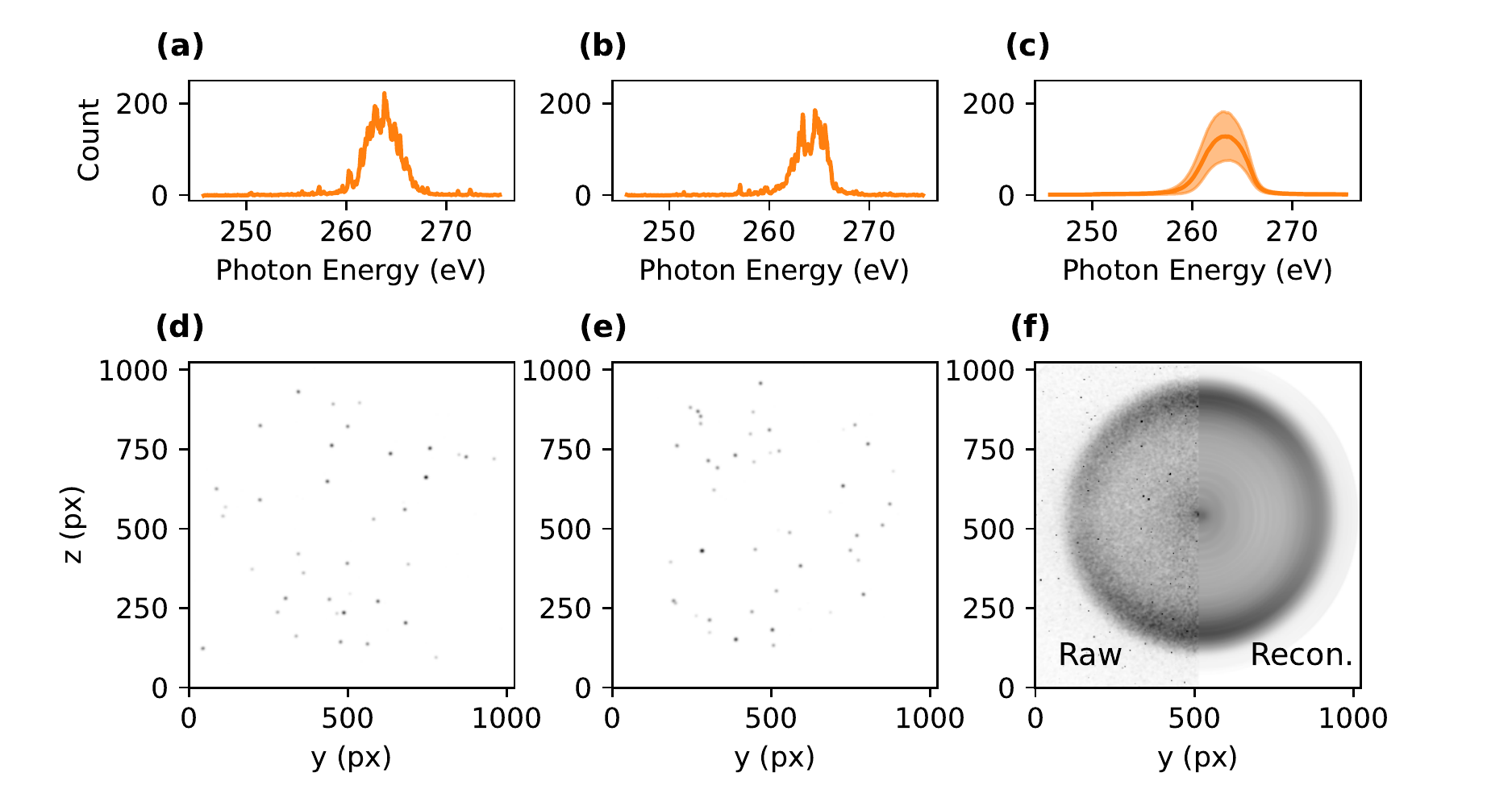}
    \caption{Representative single-shot data and average spectra. 
    \textbf{(a,b)} Two single-shot photon spectra after a global background subtraction, with
    \textbf{(d,e)} showing their corresponding VMI images. For visualization only, we have blurred the raw images with a 3px-wide (standard deviation) gaussian kernel.
    \textbf{(c)} Average photon spectrum, with the shade representing the standard deviation over all FEL shots.
    \textbf{(f)} Average VMI image, where the left half is the average raw image, and the right half is the reconstruction of average image (see main text).}
    \label{fig:single-shot}
\end{figure}

We demonstrate this reconstruction method for a dataset collected from soft x-ray ionization of argon above the $2p$ ionization threshold~\cite{heimann1987shake,avaldi1994near,jurvansuu2001inherent}.
The experiment is conducted using the CAMP instrument~\cite{erk2018camp} at beamline BL1 of the FLASH %
~\cite{feldhaus2010flash}. 
Argon gas is introduced \textit{via} supersonic expansion to produce a continuous molecular beam. 
Following collimation through two skimmers and an aperture, the beam is intercepted by focused x-ray pulses produced by the FEL in the interaction region of the spectrometer. %
The velocities of the photoelectrons are mapped to a position-sensitive microchannel plate/phosphor screen detector and recorded with a CMOS camera at $10$~Hz.  

The pulses have an average bandwidth of $5$~eV full-width at half maximum~(FHWM) at $264$~eV central photon energy. 
The estimated pulse duration is $100 - 150$~fs~(FWHM) with a mean pulse energy of $6~\mu$J/pulse at the sample. 
The incident photon spectrum is recorded on a shot-to-shot basis using an upstream variable line spacing~(VLS) spectrometer~\cite{brenner2011first}. 
The zeroth order beam from the grating is delivered to the interaction point, while the first order is collected by a detector to image the spectrum. 

\begin{figure}[hbtp]
    \centering
    \includegraphics[width=\linewidth]{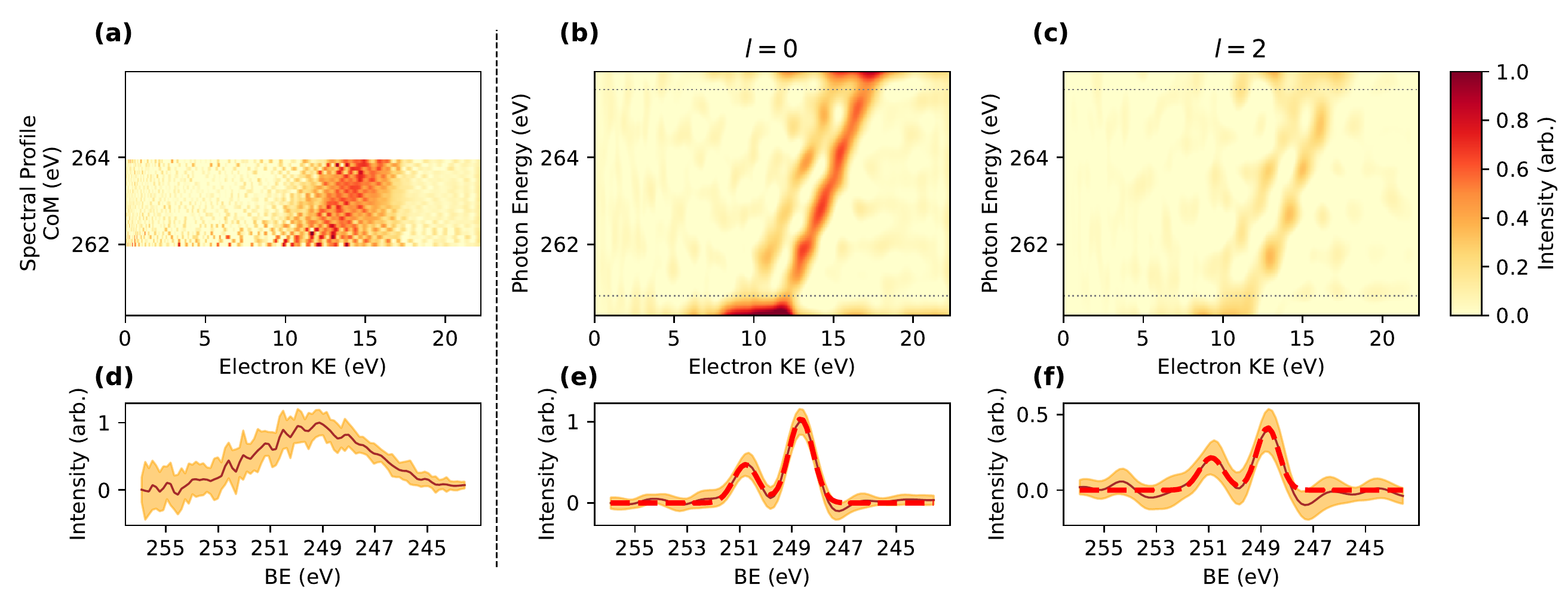}
    \caption{Comparison between the conventional binning-and-averaging method (a)(d) and our method (b)(c)(e)(f).  
    \textbf{(a)} Energy distribution obtained with pBasex applied to the cumulative VMI image, with preliminary grouping of shots by the centre-of-mass~(CoM) of spectral profile. 
    \textbf{(b-c)} Photon-energy resolved argon $2p$ photoelectron kinetic energy~(KE) spectra of the $l=0$ and $l=2$ Legendre polynomial components of the angular distribution, both being normalized to the maximum intensity in $l=0$. %
    The grey dotted lines delineate the photon energy range for (e-f).
    \textbf{(d-f)} Binding energy~(BE) spectra of (a-c) extracted by shifting the KE spectra at each photon energy, see main text. 
    The brown solid lines are the mean over the photon energy range, and the shaded regions represent the standard deviation. 
    The red dashed lines in (e-f) are the sum of fitted Gaussians.}
    \label{fig*:2dmaps}
\end{figure}

This measurement is performed with linearly polarized light from the FEL, which implies that the resulting electron momentum distribution has reflection symmetry about the $xy$ plane.
In this case, we can limit the sum in Eqn.~(\ref{eq:expand-on-basis}) to only even-order Legendre
polynomials~\cite{yang_theory_1974}. 
The dataset consists of $2.8\times 10^4$ FEL shots, and the representative single-shot raw VMI images and photon spectra are shown in Fig.~\ref{fig:single-shot}, where the average VMI image and photon spectrum are also shown.
The single-shot VMI images are quite sparse, containing 40 electrons in each image on average.

From the minimization of Eqn.~(\ref{eqn:obj}), we obtain the most probable parameters $C_{lk,w}$ given the dataset, which represents the fitted $\chi(\bm{p},\omega)$. 
Using Eqn.~(\ref{eqn:B_iq}), the average image is reconstructed by averaging the expected intensity $\hat{B}_{i,q}$ across all FEL shots. 
As shown in Fig.~\ref{fig:single-shot}~(f), the reconstructed average image agrees well with the average of raw images except being less noisy.
The reconstructed average image, nevertheless, is the projected momentum distribution of the mean photon spectrum and can hardly show the underlying fine structure.
With the same set of most-probable parameters $C_{lk,w}$, a better way to reveal the fine structure is to reconstruct the KE spectrum for each $l$-th Legendre polynomial component and each photon energy $\omega$, %
which is discretized as
\begin{equation}\label{eqn:sigma}
    I_{l}(E_r, \omega_w) = \sum_k C_{lk,w}f_k(\sqrt{\frac{E_r}{\alpha}})\sqrt{\frac{E_r}{4\alpha^3}}~,
\end{equation}
with $E_r=\alpha r^2$ denotes the KE at radial grid point $r$. 
The ionization from argon $2p$-shell is a single-photon process, so we visualize %
$I_0(E,\omega)$ and $I_2(E,\omega)$, in Fig.~\ref{fig*:2dmaps}~(b)(c) respectively.

We observe two dispersive features
that correspond to the spin-orbit split cationic states of argon, $^{2}P_{1/2}$ and $^{2}P_{3/2}$~\cite{jurvansuu2001inherent}. 
For each photon energy in Fig.~\ref{fig*:2dmaps}~(b), we shift the reconstructed KE spectrum to binding energy~(BE) according to $\mathrm{BE}=\hbar\omega-\mathrm{KE}$, which is shown in Fig.~\ref{fig*:2dmaps}~(e).
The spin-orbit splitting is well resolved with our approach. 
We use a previously reported measurement~\cite{jurvansuu2001inherent} of this splitting, $2.12$~eV, to calibrate the VMI and VLS energy axes. 
With this calibration, we extract a line-width of $1.1$~eV~FWHM from Fig.~\ref{fig*:2dmaps}~(e), which is resolved beyond the width of the averaged FEL pulse spectrum, $5$~eV FWHM. 
Furthermore, the extracted line-width is below the average bandwidth of a single pulse $3.5$~eV FWHM.
Thus, with the application of spectral domain ghost imaging in this XFEL experiment, we are able to achieve sub-bandwidth resolution. 

In Fig.~\ref{fig*:2dmaps}~(c)~and~(f) we plot the projection of the 3D momentum distribution onto the second-order Legendre polynomial $I_2(E,\omega)$, which is related to the photoemission anisotropy parameter, $\beta_2$~\cite{cooper_angular_1968}.
We extract a value of $0.46\pm0.05$ and $0.39\pm0.05$ for the anisotropy parameter for the $^{2}P_{1/2}$ and $^{2}P_{3/2}$ ionic states, respectively.
These values are averaged over a range of photon energies from $261$ to $265$~eV, which is in good agreement with previous measurements in Ref.~\cite{Avaldi1994Ar2p} and calculations in Ref.~\cite{Lindle_PRA_Angleresolved1988}. 
The results in Fig.~\ref{fig*:2dmaps}~(b)(c)(e)(f) are obtained with a set of $N_K=125$ radial basis functions and uniform pixel weight $W$.

In contrast, directly applying pBasex on the cumulative image results in a single spectral feature with a width of $5.4$~eV FWHM. %
Moreover, grouping the FEL shots by the centre-of-mass of photon spectrum does not resolve the two spin-orbit features, and the corresponding average binding energy spectrum has a width of $5.1$~eV (FWHM), as shown in Fig.~\ref{fig*:2dmaps}~(a)~and~(d).
In order to obtain the resolution observed in Fig.~\ref{fig*:2dmaps}~(b)~and~(c), we need to explicitly consider the correlations between the single-shot spectra and the VMI images.

\section{Discussion}\label{sec:Discussion}

\subsection{Resolution achieved in our demonstration experiment}
To characterize the resolution of this technique, it is useful to define the correlation length of the spectral measurement.
The correlation length is a measure of how strongly coupled the intensity fluctuations are between neighboring pixels in the photon spectrum. 
Here, we define this metric $\delta_A$ to be the average distance between two frequencies where the Pearson's correlation coefficient drops to $1/e$~\cite{driver_attosecond_2020}. 
The correlation between spectrometer pixels arises from two sources: the resolution of the photon spectrometer, and the intrinsic variation due to spectral fluctuations of the source.
In our experiment, the correlation length is measured to be $\delta_A=22$~VLS pixels, corresponding to $0.99$~eV, as shown in Fig.~S1(a) of the supplementary material.
This is much larger than the instrumental resolution of the photon spectrometer~\cite{brenner2011first}, and thus, we conclude that the dominant contribution to the correlation length is the intrinsic correlation of the FEL source. 

The average width of the photoemission features in the BE spectrum in Fig.~\ref{fig*:2dmaps}~(e) is determined to be $\sigma_\mathrm{BE}=0.48~\mathrm{eV}$ in standard deviation. 
This width is the result of the finite VMI energy resolution and the correlation in the spectral measurement, which approximately sum in quadrature $\sigma^2_\mathrm{BE} \approx \sigma^2_\mathrm{VMI} + \sigma^2_{\phi\mathrm{Corr}}$. 
The contribution from spectral correlation, $\sigma_{\phi\mathrm{Corr}}$, is not necessarily the correlation length $\delta_A$, %
but the two quantities are related by a constant of proportionality. 
As described in the supplemental material, and shown in Fig.~S1(b),  $\sigma_{\phi\mathrm{Corr}}^2$ is estimated to be $0.11~\mathrm{eV}^2$ based on the measured $\delta_A=22$~px. 
Having accounted for $\sigma_{\phi\mathrm{Corr}}^2$, we find the remaining contribution to $\sigma^2_\mathrm{BE}$ to be $\sigma_\mathrm{VMI}^2=0.10~\mathrm{eV}^2$.
This corresponds to a VMI energy resolution of 2.3\% at $\mathrm{KE}=15~\mathrm{eV}$, which is consistent with the reported value of 2\% for this instrument~\cite{erk2018camp}.
We note that the average bandwidth of incident x-ray pulses is not a limiting factor for the resolution $\sigma_\mathrm{BE}$ of our technique. 
Instead, it is determined by the kinetic energy resolution of the VMI spectrometer and the correlation length of the photon spectrum measurement.

\subsection{Extension of SDGI}\label{sec:asSDGI}

Spectral domain ghost imaging, in general, exploits spectral fluctuations of the incident source to capture the sample response. 
The key development in the present work is incorporating the native projection of VMI into the ghost imaging model.
This method unravels the projection of different Newton spheres, while simultaneously correlating these unravelled shells with the incident photon spectra.
More specifically, our approach models the photoproduct momentum space with a reduced dimensional representation~(similar to the pBasex approach) and inverts the tensor product of spectral integration and Abel transform in a single step.
Such an approach is a unique way to extend the scope of SDGI. 
Although under ideal circumstances it is equivalent to conducting the two regression steps sequentially, the single-step approach has several advantages over either sequential approach.

In comparison to the sequential application of SDGI followed by Abel inversion, the single-step approach represents the momentum distribution with a polar basis set, reducing the dimensionality of the problem and coordinating the comprehensive behavior of the momentum distribution over multiple pixels. 
When applying SDGI directly to the 2D VMI data, incorporating the smoothness term that penalizes high spatial frequencies in momentum space is inefficient and in most cases intractable. 
This is because the dimension of the image space is too large. 
Conceding this smoothness over momentum space, we can implement a sequential approach by applying SDGI to each VMI pixel independently and subsequently applying pBasex to the result. 
This approach remains efficient, but it is plagued by compromised performance in noise handling, as shown in the supplementary material Fig.~S2. %
The key difference between this sequential approach and the single-step is that the regression is performed independently for each pixel in the first of the two sequential steps.

Another feasible sequential approach is to apply single-shot Abel inversion followed by SDGI. 
Although the regularization terms can remain in the same form as in the single-step approach, in this approach the Gaussian log-likelihood in the objective function~(Eqn.~\ref{eqn:obj}) is replaced by,
\begin{equation}\label{eq:alt-h0}
    h_{0,\mathrm{p+S}}(C; A,G,B) = \sum_{i} \left|\sum_{w}A_{i,w}C_{lk,w}- \sum_q\left(G^+\right)_{q,lk} B_{i,q}\right|^2~,
\end{equation}
where $G^+ = (GWG^T)^{-1}GW$ is the pseudo-inverse of the Abel transform, which is conducted separately for each photon energy bin. 

In our demonstration experiment, both sequential approaches show inferior performance compared to the single-step approach, which is illustrated in the comparison in the supplementary material.
Under ideal circumstances, when regularization becomes unnecessary to suppress experimental noise, %
the optimal point of the original objective function in Eqn.~(\ref{eqn:obj}) is
\begin{subequations}
\begin{eqnarray}\label{eqn:unreg-solution}
    C_{lk,w}^{(\lambda=0)} &=& \sum_{i,q} \left((G\otimes A)^+\right)_{wlk,iq} B_{i,q}\\ 
    &=& \sum_{i,q} \left(A^+\right)_{w,i}\left(G^+\right)_{q,lk} B_{i,q}~,
    \label{eqn:unreg-solution2}
\end{eqnarray}
\end{subequations}
where $A^+ = (A^TA)^{-1}A^T$, and in going from Eqn.~(\ref{eqn:unreg-solution})~to~(\ref{eqn:unreg-solution2}) we have changed the order of operations between the pseudo-inverse and tensor-product of $A$ and $G$.
Equation~(\ref{eqn:unreg-solution}) describes the single-step approach and Eqn.~(\ref{eqn:unreg-solution2}) denotes both the sequential approaches described above.
Thus it can be seen that in the absence of regularization, both sequential approaches are equivalent to the single-step method.
The regularization in Eqn.~(\ref{eqn:obj}) necessitates the computationally heavy step of inverting an $N_\omega N_LN_K$-dimensional matrix, which is not true for the sequential approaches. 
However, as described above, the single-step outperforms the sequential approaches by virtue of being more robust to noise.
Similar results between the single-step and sequential approaches may be achieved in other measurements, and the precise choice of method depends critically on the properties of the measurement; e.g. signal-to-noise ratio, size of the dataset, and number of parameters to fit.

\subsection{Connection to Covariance Approach}

We note that there are a number of mathematical techniques to extract correlations in large datasets.
While the method presented in this work relies on linear regression, covariance analysis is another common method for extracting correlations~\cite{frasinski1989covariance,frasinski2013dynamics,frasinski_covariance_2016,allum_loc_2022,amiot_supercontinuum_2018}.
The photoproduct yield can be correlated with other intrinsic or extrinsic measurements, to extract signal from noisy data. %
Both regression and covariance have been employed in ghost imaging experiments in the spectral domain, and these two techniques are closely related.

The connection is shown by regressing the mean-subtracted VMI images $\Delta B_i \equiv B_i-\langle B\rangle$ on the mean-subtracted spectra $\Delta A_i\equiv A_i-\langle A\rangle$, with $\langle\cdot\rangle$ denoting the average across all shots.  
Given the standard definition of sample covariance 
$\cov[X,Y]=\Delta X^T \Delta Y / (N_\mathrm{shot}-1)$, %
the unregularized spectral regression of $\Delta B$ on $\Delta A$ gives
\begin{equation}
    (\Delta A)^+\Delta B = (\Delta A^T \Delta A)^{-1} (\Delta A^T \Delta B) = \cov[A,A]^{-1}\cov[A,B]~,
    \label{eqn:cov-analysis}
\end{equation}
where the right hand equality demonstrates this is identical to applying the inverse of the photon autocovariance matrix to the photon-electron covariance matrix. 

The transform from the projected momentum distribution to the coefficients $C$ is a linear operation in the electron momentum space, 
which is in tensor product with the operations in the photon energy space. 
Therefore the aforementioned single-step and sequential regression approaches are all applicable to the mean-subtracted data. 
Comparing to the original approach elaborated in Sec.~\ref{sec:method}, regressing the mean-subtracted data is more robust to static background in both the VMI images and photon spectra, but it may suffer from the loss of information due to subtraction of the mean, especially when the spectral fluctuation is limited. 
Regardless whether the mean is subtracted or not, one can further constrain the parameters with additional prior knowledge about the sample, to solve the most probable parameters in a space with dimension lower than $N_\omega N_LN_K$. 
Although related, these further restricted approaches require more assumptions and are less generalisable than the main approach we present in this work. 
We provide an example in the supplementary material.

\subsection{Applicable Apparatus Configurations}\label{sec:configs}
As for standard inverse Abel transform procedures, a prerequisite of our method is the cylindrical symmetry about any axis $z$ that is perpendicular to the projection axis of the VMI $x$, for which the coordinate system shown in Fig.~\ref{fig:measurement} is not the only configuration. 
Here we describe two other configurations without exhausting all possible cases. 
For circularly polarized x-rays, such as in~\cite{Ilchen_Site_2021}, the layout in Fig.~\ref{fig:measurement} is still applicable except that the symmetry axis $z$ is along the axis of beam propagation. 
For a co-axial VMI~\cite{li_co-axial_2018} with linearly polarized x-rays propagating along the VMI axis $x$, $z$ falls along the x-ray polarization axis.

\section{Conclusion and Outlook}

We present a regression approach which can achieve VMI measurements at resolution better than the inherent bandwidth of a noisy photon source, by simultaneously reconstructing the spectral response and performing the inverse Abel transform. 
Our approach demonstrates a clear advantage over the conventional binning-and-averaging method.
In our experimental demonstration on the photoionization of argon, the retrieved linewidth is dominated by the resolution limit prescribed by the VMI resolution and the measured spectral correlation. 
We anticipate that the outlined approach will be of great use in the emerging field of time-resolved inner-shell photoelectron spectroscopy at FELs, which promises to interrogate photoinduced nuclear and electronic dynamics in a site-selective manner.

We demonstrate the connection between covariance and regression analysis and %
we relate and compare our single-step approach to other sequential approaches. 
Our method extends the scope of SDGI by adopting a reduced dimensional representation for the properties of the photoproducts, and it can be directly applied to any measurement where VMI images are recorded in coincidence with the incident photon spectrum.
This makes the technique broadly applicable to many different light sources. 

\section{Data Availability}
The data that support the findings of this study are available from the authors upon reasonable request. 
An implementation of the proposed approach can be found in our custom package developed for general SDGI~\cite{Wang_Spook_-_A_2022}.

\section{Author Contributions}
T.C.D., J.P.C and R.F. initiated the research. J.W., T.C.D. and J.P.C conceived the method, which is implemented by J.W.  The experiment was carried out by C.C.P., C.P., G.B., S.D., A.T.N., S.K., and B.E.~. All authors contributed to collection or analysis of the experimental data and writing of the manuscript.

\section{Acknowledgement}
We acknowledge the Max Planck Society for funding the development and the initial operation of the CAMP end-station within the Max Planck Advanced Study Group at CFEL and for providing this equipment for CAMP@FLASH. 
The installation of CAMP@FLASH was partially funded by the BMBF grants 05K10KT2, 05K13KT2, 05K16KT3 and 05K10KTB from FSP-302.
The analysis work was supported by the U.S. Department of Energy (DOE), Office of Science, Office of Basic Energy Sciences (BES), Chemical Sciences, Geosciences, and Biosciences Division (CSGB). 
We acknowledge DESY (Hamburg, Germany), a member of the Helmholtz Association HGF, for the provision of experimental facilities. 
Parts of this research were carried out at FLASH and the Maxwell computational resources operated at Deutsches Elektronen-Synchrotron DESY, Hamburg, Germany. 
We thank Jordan O'Neal and Alice E. Green for fruitful discussions.

\bibliographystyle{apsrev4-1}
\bibliography{main.bib, referencesJPC.bib}

\end{document}


\title{Supplementary Material for "Ghost Imaging Approach to Photon Energy Resolved Velocity Map Imaging"}

\author{Jun Wang \textit{et al}}
\maketitle

\section{Correlation Length of XFEL Pulses and Resolution in the Retrieved Electron Spectra}
\begin{figure}[hbtp]
    \centering
    \includegraphics[width=0.85\linewidth]{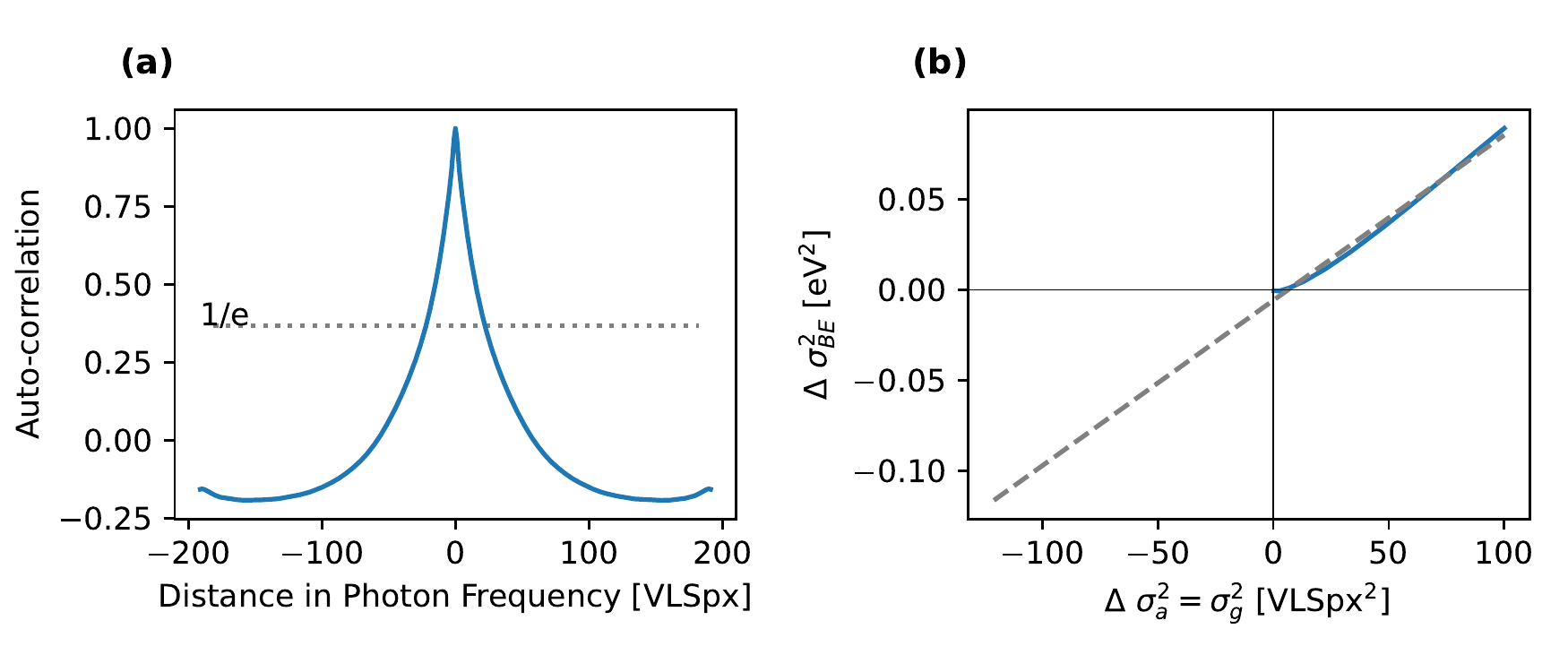}
    \caption{Spectral correlation and its effect on binding energy resolution. 
    \textbf{(a)} Average autocorrelation function of XFEL Pulses used in the argon experiment. The gray dotted line delineates $1/e$ on vertical axis.
    \textbf{(b)} Dependence of increment of the average linewidth squared $\Delta \sigma_\mathrm{BE}^2$ on the Gaussian-blurring. The gray dashed line is the linear extrapolation to $\Delta\sigma_a^2 = -\sigma_{a0}^2$.}
    \label{suppfig:corrlen}
\end{figure}

The fluctuation of the spectral intensity of XFEL pulses has an intrinsic correlation between neighboring frequencies. 
The correlation between intensity at two frequencies can be quantified with Pearson's correlation coefficient
\begin{equation}
    C_\mathrm{P}\left(A(\omega_1), A(\omega_2)\right) = \frac{\langle \Delta A(\omega_1) \Delta A(\omega_2) \rangle}{\sqrt{\langle \Delta A(\omega_1)^2\rangle \langle \Delta A(\omega_2)^2\rangle}}~,
\end{equation}
where $\Delta A(\omega) \equiv A(\omega) - \langle A(\omega)\rangle$ is the fluctuation of the spectral intensity about the mean value.
To extract a single correlation length, we average over a range of frequencies, $\omega_L<\omega<\omega_U$, where the mean photon spectrum is above $1/e^2$ of the maximum to obtain the average autocorrelation function
\begin{equation}
    F(\Omega) \equiv \int_{\omega_L}^{\omega_U} C_\mathrm{P}\left(A(\omega+\frac{\Omega}{2}),  A(\omega -\frac{\Omega}{2})\right) \frac{d\omega}{\omega_U-\omega_L}~,
\end{equation}
and the correlation length is defined to be the frequency $\Omega_c$ that the function falls to $F(\Omega_c) = 1/e$. 
The autocorrelation function of the $2.8\times 10^4$ XFEL pulses used in our demonstration experiment is shown in Fig.~\ref{suppfig:corrlen}(a), and the correlation length is determined to be $\delta_A=22$ VLS pixels, while the instrumental resolution is $\sim 1$~px~\cite{brenner2011first}.

The correlation length $\delta_A$ is not exactly the same as $\sigma^2_{\phi\mathrm{Corr}}$ for at least two reasons.
First, the autocorrelation function is not exactly the photon spectrum, but the normalized auto-covariance of the fluctuating spectrum.
Therefore, if the point spread in the photon spectrum is modelled with a gaussian of width $\sigma_{a0}$, the autocorrelation function is also expected to be a gaussian with standard deviation $\sqrt{2}\sigma_{a0}$, and so at the $1/e$ of the maximum, $\delta_A=2\sigma_{a0}$, thus we estimate $\sigma_{a0}=11$px .
Secondly, averaging over $\omega$ in obtaining the binding energy spectrum may have alleviated the blurring due to the finite $\delta_A$.
Both effects maintain the linear dependence of $\sigma^2_{\phi\mathrm{Corr}}$ on $\delta_A$, but the equality may not be exact. 

Returning to the discussion in Section~IV.A, recall that the finite resolution of the electron binding energy spectrum is given by, $\sigma^2_\mathrm{BE} \approx \sigma^2_\mathrm{VMI} + \sigma^2_{\phi\mathrm{Corr}}$.
Rather than estimating $\sigma_{\phi\mathrm{Corr}}$ directly from a measurement, it is easier to increase $\sigma^2_{\phi\mathrm{Corr}}$ by blurring the photon spectra, and extrapolate the behavior of this curve.
To estimate $\sigma^2_{\phi\mathrm{Corr}}$ in our experiment, we blur the raw VLS spectra with a gaussian kernel of varying width $\sigma_g$. 
At each $\sigma_{g}$ we repeat the single-step regression to obtain the binding energy spectrum, and we fit the two features to a double-gaussian model, obtaining an average linewidth $\sigma_{BE}$ of the two spin-orbit states as a function of $\sigma_g^2$. 
The double-gaussian model is constrained to have the same standard deviation for both peaks, which is reasonable in this case because the intrinsic linewidths of the two states only differ by $1$meV~\cite{jurvansuu2001inherent}.

Convolved with the gaussian kernel, the gaussian point spread model of the photon spectrum remains gaussian, but broadened to $\sigma_a = \sqrt{\sigma_{a0}^2+\sigma_g^2}$. 
Given the slope shown in Fig.~\ref{suppfig:corrlen}(b) , we can extrapolate to $\sigma_a^2 = \sigma_{a0}^2 + \sigma_{g}^2  =0$, 
and we estimate that $\sigma_\mathrm{BE}^2$ would decrease by $0.11~\mathrm{eV}^2$ if $\delta_A$ were 0, and thus we take $\sigma_{\phi\mathrm{Corr}}^2=0.11~\mathrm{eV}^2$ as our estimate, as stated in the main text.

\section{Comparison to Sequential Approaches}
Besides the single-step approach presented in the main text, there are two sequential approaches to apply spectral-domain ghost imaging~(SDGI) and Abel inversion with pBasex to the data.
One method involves independently conducting SDGI on each pixel of the VMI detector followed by applying pBasex to the SDGI reconstruction.
The other method applies pBasex to single-shot VMI images followed by SDGI on the single-shot coefficients.
In the first sequential method, the smoothness regularization over the image space is inefficient, or even intractable to implement in some cases, because a VMI image typically has $(4\times 10^2)^2\approx 2\times 10^5$ pixels in a quadrant, which amounts to a forbidding dimension of $(2\times 10^5)^2/2=2\times 10^{10}$ for the smoothness operator. 

\begin{figure}[htbp]
    \centering
    \includegraphics[width=0.6\linewidth]{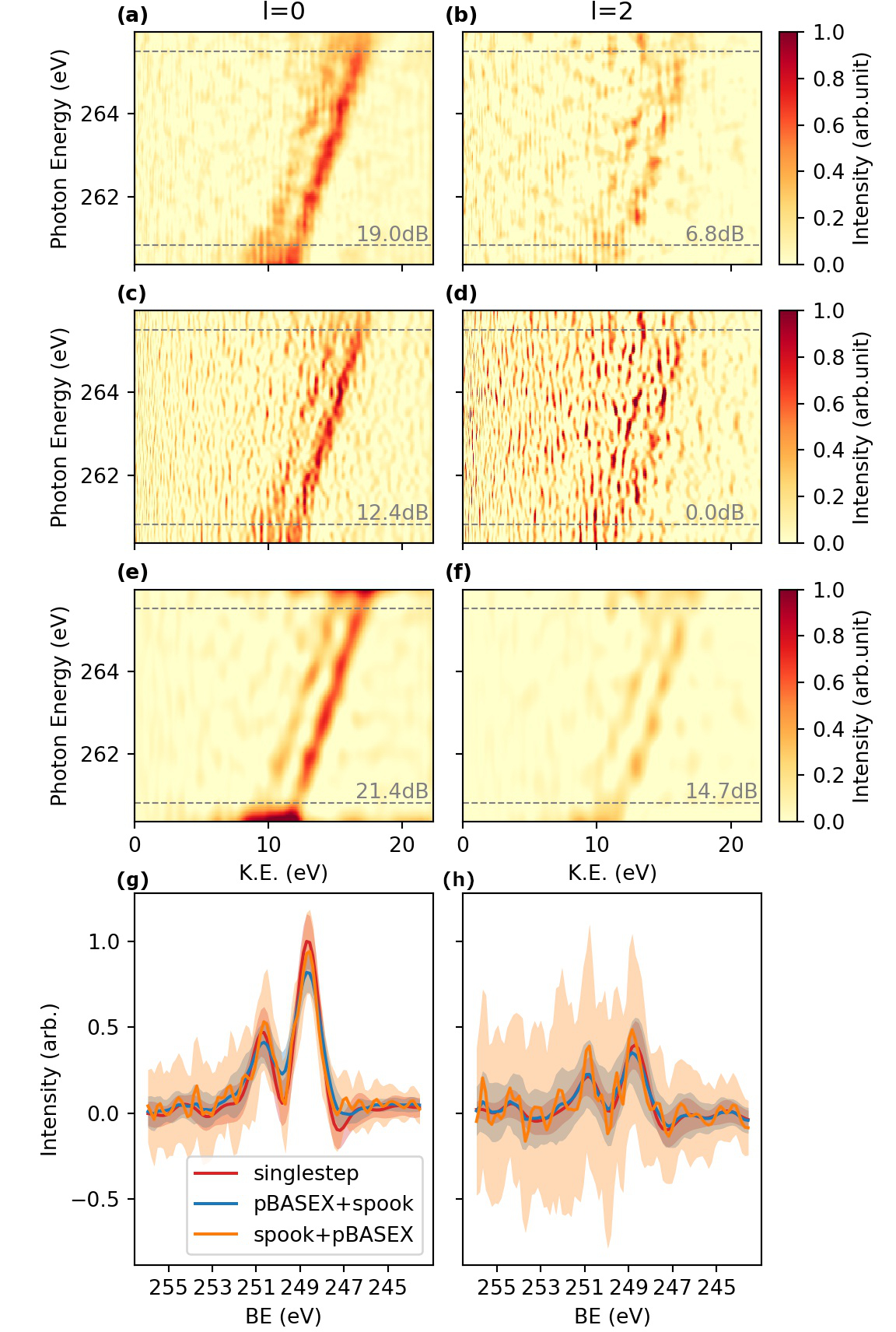}
    \caption{Comparison between the two-step approaches and the single-step approach. 
    (a-b) pBasex then SDGI; (c-d) SDGI then pBasex; (e-f) single-step, the same as Fig.4(b-c) in the main text. 
    For each approach, the photoelectron spectra of the $l=0$ and $l=2$ Legendre polynomial components are shown separately but both normalized by the same factor. 
    The peak-signal-to-noise-ratio between the gray dashed lines is shown in gray at the lower right in each panel. (g-h) the corresponding binding energy spectra of the ones in (a-f). The transformation from $I(\mathrm{KE}, \omega)$ to $I(\mathrm{BE})$ is explained in the main text.}
    \label{suppfig:2stepvs1step}
\end{figure}

Both sequential approaches should perform well given sufficient number of hits per image, sufficient number of shots and sufficient computational power. 
Given the same realistic dataset, however, they are less robust to noise than the single-step approach we describe in the main text.
As shown in Fig.~\ref{suppfig:2stepvs1step}~(a-d), both sequential approaches are able to resolve the two spin-orbit states, but the signal-to-noise ratio is lower than the results obtained with the single-step approach in Fig.~\ref{suppfig:2stepvs1step}~(e-f).
The is particularly true for the $l=2$ Legendre polynomial components.
We quantify this result using the peak-signal-to-noise ratio~(PSNR), referencing to the fitted result with a double-gaussian model: 
\begin{equation}\label{suppeq:psnr}
    \mathrm{PSNR}\left(I(\mathrm{KE},\omega), I_\mathrm{ref}(\mathrm{KE},\omega)\right) = 20\log_{10}\left(\frac{\max_{\mathrm{KE},\omega} I_\mathrm{ref}}{\sqrt{\langle (I-I_\mathrm{ref})^2\rangle}}\right)~,
\end{equation}
where 
$I(\mathrm{KE},\omega)$ denotes a spectrum of interest, 
$I_\mathrm{ref}(\mathrm{KE},\omega)$ denotes the reference spectrum, and $\langle\cdot\rangle$ denotes the average over $(\mathrm{KE},\omega)$ space. 
In our case, the reference $I_\mathrm{ref}$ is obtained by fitting $I(\mathrm{KE},\omega)$ to a model containing two photoelectron features with Gaussian lineshape:
\begin{equation}
    I_\mathrm{model}(\mathrm{KE},\omega; A,\mu,\sigma) = \sum_{v=1}^2 A_v \exp\left(\frac{-(\mathrm{KE}-\hbar\omega-\mu_v)^2}{2\sigma_v^2}\right)~.
\end{equation}

It is interesting to note that, in the pBasex+SDGI sequential approach, applying the Gaussian-likelihood SDGI to single-shot pBasex results is robust to reconstruction artifacts caused by the sparsity of hits in single-shot images.
The coefficients resulting from single-shot Abel inversion $G^+B$ only come into the SDGI objective function through the discrepancy $h_{0,p+S}$ defined in the main text Eqn.~(9), 
and that is again quadratic in $C$, so $G^+B$ only involves in the optimization through the linear term in $C$, whose coefficient is $\sum_{i,q} A_{i,w} \left(G^+\right)_{q,lk} B_{i,q}$.
As the shot index $i$ is summed over, we see that the Abel inversion is equivalently applied to the average of images weighted by spectral intensity.
In other words, what appeared to be a single-shot Abel inversion is the Abel inversion on the averaged image weighted by $A_w$, conducted for each photon energy bin $\omega_w$ separately. 
In short, due to the linearity of Abel inversion and the fact that Gaussian-likelihood effectively contracts shot index, single-shot Abel inversion is not detrimental.
This explains why in Fig.~\ref{suppfig:2stepvs1step} , the pBasex+SDGI sequential approach outperforms the SDGI+pBasex approach.

\section{Hyperparameter Tuning}
The result of regularized regression depends on the hyperparameters that control the regularization terms. 
In the previous works~\cite{Li_JPB2021, DriverPCCP, Klein_HighResolution_2022}, where regularized regression spectral-domain ghost imaging was applied, the hyperparameters are either tuned in simulation or set empirically. 
In this work, we use $k$-fold cross validation~\cite{Hastie_CV_2009} to evaluate the performance of models with different hyperparameters.
The metric employed to score the models is the mean $\mathrm{L}_2$ residue averaged over the validation sets, as explained below.
We inspect the resulting $\omega$-dependent energy spectrum from the hyperparameters that well-perform in terms of the metric. 
The hyperparameter is chosen based on the joint consideration of the residue's dependence on hyperparameters and the qualitative behavior of the resulting spectrum.

The $k$-fold cross validation procedure is conducted as follows. 
We randomly shuffle and split the whole dataset into $k$ sets. 
Holding out one as the validation set, we solve the optimization problem with the rest of the dataset~(i.e. the training set), the objective function being specified by Eqn.~(6) in the main text. 
At each hyperparameter value $(\lambda_\mathrm{sp}, \lambda_\mathrm{sm,\omega}, \lambda_\mathrm{sm,r})$, we permute through all $k$ possible validation sets and obtain $k$ fitted models, each of which is evaluated on their corresponding validation set, and the average residue is the performance at this hyperparameter value. 
Specifically, the mean $\mathrm{L}_2$ residue of a model represented by coefficients $C$ on a validation set $(A^{(v)}, B^{(v)})$ is
\begin{equation}
    h_{0ms} = \sqrt{\frac{1}{N^{(v)}}h_0(C;A^{(v)},G,B^{(v)})}~,
\end{equation}
where $h_0$ is squared-error function defined in Eqn.~(5) in the main text,  $N^{(v)}$ is the number of shots in the validation set. 

Note that the training set is normalized prior to optimization, which helps reduce the effective magnitude of hyperparameters to around unity, otherwise the magnitude of hyperparameters that have effects on the outcome may be hard to search. 
Specifically, we normalize the photon spectrum $A$ and the projections of basis $G$ such that $\sum_{i,w}A_{i,w}=N_\omega,~ \sum_{kl,q} G_{kl,q}^2=N_q$ . 
The outcome optimal $C$ has been scaled back to the original scale before scoring. 

\begin{figure}[htbp]
    \centering
    \includegraphics[width=0.9\linewidth]{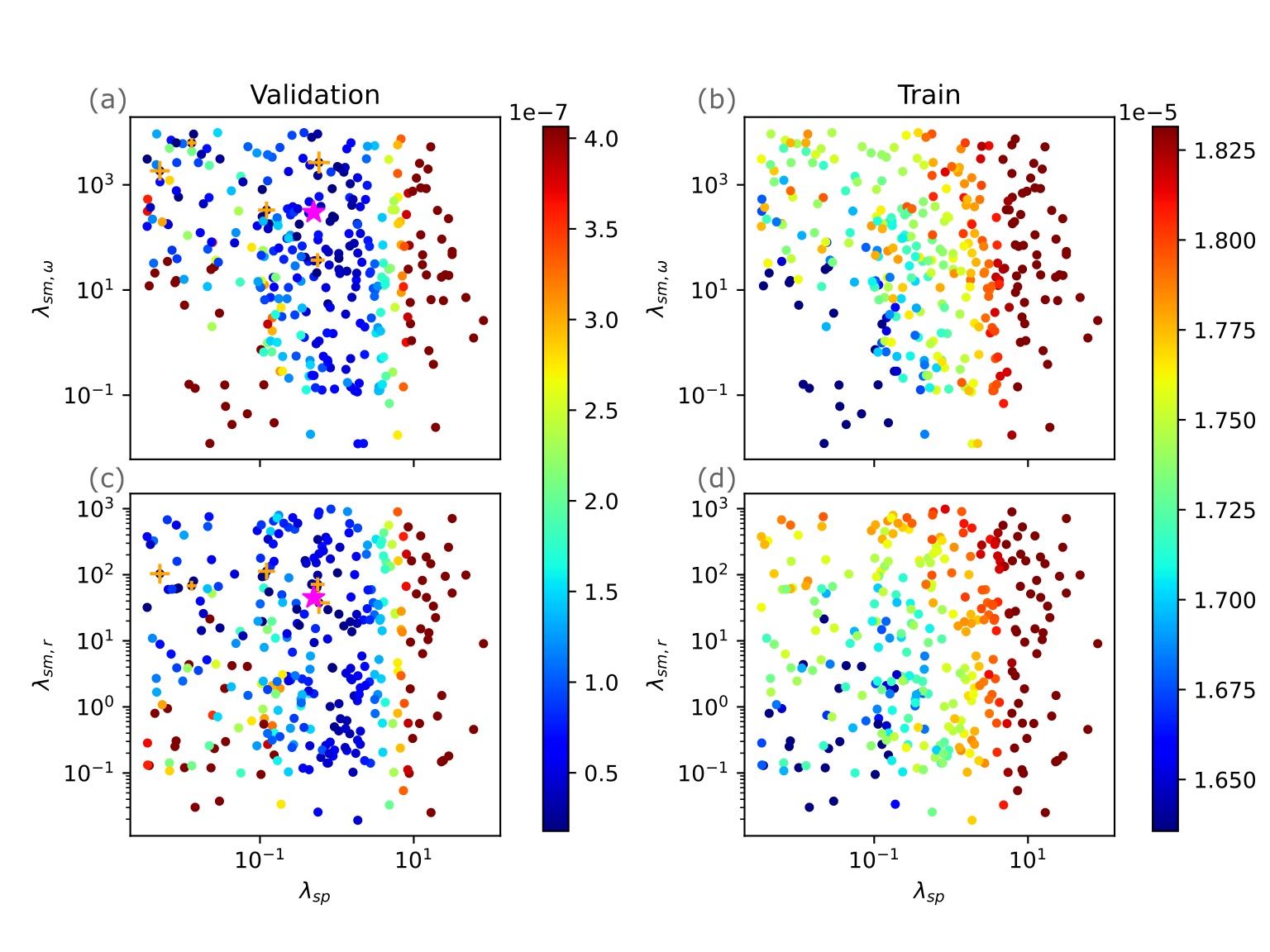}
    \caption{Hyperparameter tuning with $10$-fold cross validation. The dependence of mean $\mathrm{L}_2$ residue on hyperparameters $(\lambda_\mathrm{sp},\lambda_\mathrm{sm,\omega}, \lambda_\mathrm{sm,r})$, evaluated on (a,c) validation sets and (b,d) training sets. A global baseline of 5.5 has been subtracted prior to visualization. 
    The three dimensional space $(\lambda_\mathrm{sp},\lambda_\mathrm{sm,\omega}, \lambda_\mathrm{sm,r})$ is projected into the $(\lambda_\mathrm{sp},\lambda_\mathrm{sm,\omega})$ and $(\lambda_\mathrm{sp},\lambda_\mathrm{sm,r})$ spaces.
    The best five hyperparameters in validation metric are marked by orange crosses in (a,c), with increasing marker size indicating higher ranking.
    The magenta star marks the hyperparameter we selected.}
    \label{suppfig:crossval}
\end{figure}

We cross validate with a $k=10$ split. 
The hyperparameters are log-uniformly sampled in several patches of region.
The same residue evaluated on the training sets is expected to by and large decrease with decreasing hyperparameters. 
Thus it can serve as a sanity check and provide a reference for the level of fluctuation in the hyperparameter tuning.
We visualize the metric on both validation and training sets in Fig.~\ref{suppfig:crossval} .
The five best points in validation metric are marked by the crosses in Fig.~\ref{suppfig:crossval}~(a)~(c), and after inspecting the resulting spectra, we choose $(\lambda_\mathrm{sp},\lambda_\mathrm{sm,\omega}, \lambda_\mathrm{sm,r}) = (0.5, 300, 45)$ as marked by the star.

\section{Modeling with Binding Energy Spectrum}
In this approach, we make use of the prior knowledge that, in the range of photon energy involved in this argon experiment,
\begin{enumerate}
    \item only single-photon dispersive features are present,
    \item their cross-section is, to a good approximation, independent from photon energy.
\end{enumerate}
With these additional assumptions, we can further reduce the model parameters to the ionization intensity spectrum in binding energy $s(\mathrm{BE})$. 
In this way, the photoelectron spectrum in KE is expected to be a convolution between the spectrum $s(\mathrm{BE})$ and photon spectrum $a(\omega)$
\begin{equation}
    \hat{I}(E) = \int d\omega s(\omega-E) a(\omega)~,
\end{equation}
and photon-electron covariance in energy-energy space is expected to be
\begin{equation}\label{suppeq:CAIexp}
    F[s](\Omega,E)\equiv \cov[\hat{I}(E), A(\Omega)] = \int d\omega s(\omega-E) \cov[A(\omega), A(\Omega)]~.
\end{equation}
We invert photon autocovariance out of photon-electron covariance by minimizing the objective function in Eqn.~(\ref{suppeq:main}) subject to $s(\mathrm{BE})\geq 0$
\begin{equation}
    h(s) = \sum_{w,r} \big|\cov[A,I](\omega_w,E_r)-F[s](\omega_w,E_r)\big|^2 + \lambda_1 \sum_{b} |s_b| + \lambda_2 \sum_{b} |s''_b|^2 ~, 
    \label{suppeq:main}
\end{equation}
where $s''_b$ is the numerical second-order derivative at point $b$, and $\lambda_1$ and $\lambda_2$ are regularization hyperparameters.

\begin{figure}[hbtp]
    \centering
    \includegraphics[width=0.9\linewidth]{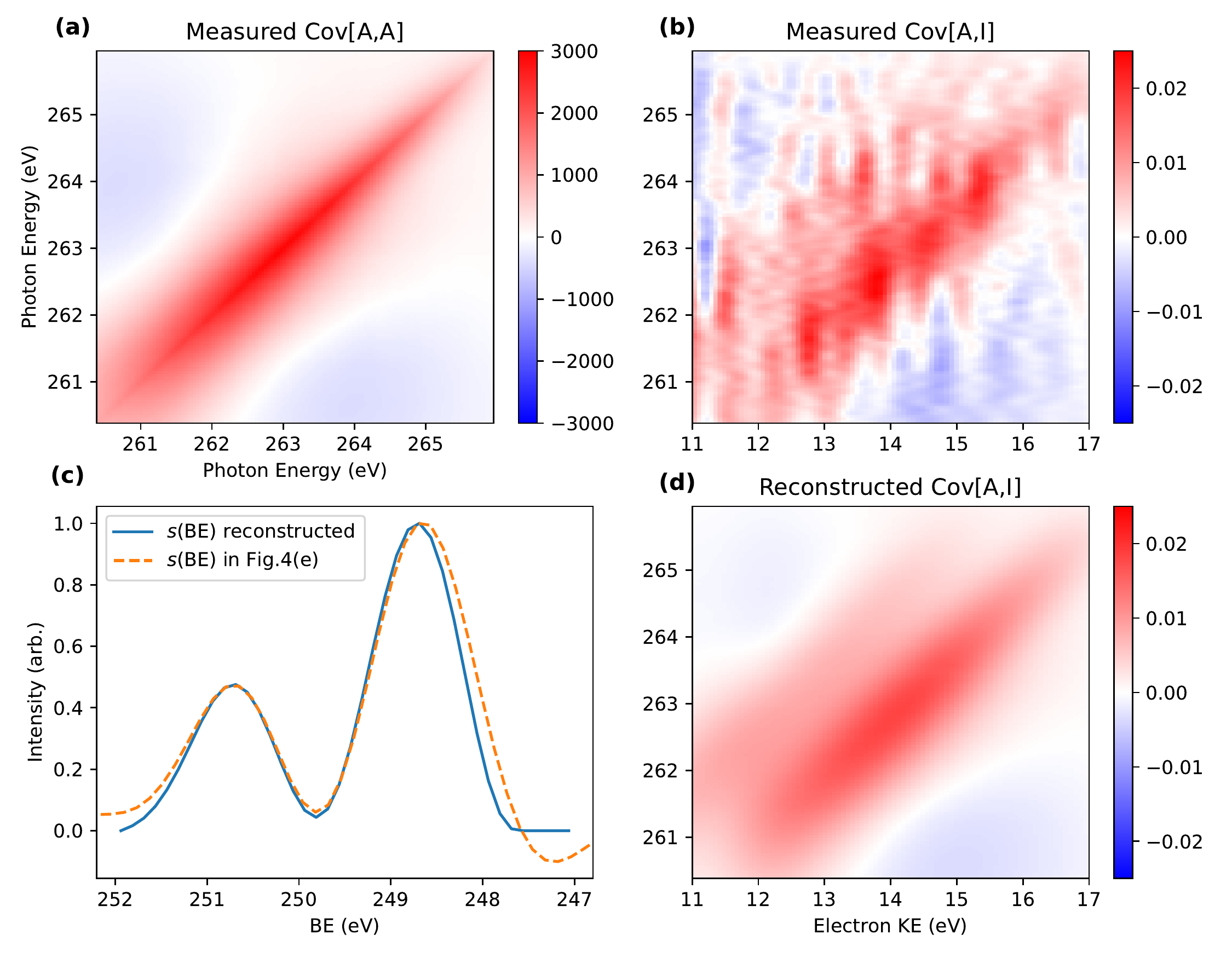}
    \caption{Covariance analysis of the argon ionization experiment. (a)\,Measured photon autocovariance map $\cov[A,A](\omega,\omega)$ and (b)\, Photon-electron covariance map $\cov[A,I](\omega,E)$. (c)\, Reconstructed $s(\mathrm{BE}$ (solid) as compared to the BE spectrum obtained in Fig.4(e) in the main text (dashed). (d)\, Reconstructed $\cov[A,I](\omega,E)$ according to Eqn.~(\ref{suppeq:CAIexp}) . }
    \label{suppfig:sBEresults}
\end{figure}

We apply the linear transforms in electron momentum space to the photon-electron-image covariance $\cov[A,B]$, obtaining the photon-electron covariance in energy-energy representation $\cov[A,I]$. 
The covariance matrices $\cov[A,I]$ and $\cov[A,A]$ are visualized in Fig.~\ref{suppfig:sBEresults} (a)(b).
The two spin-orbit features are visible in $\cov[A,I]$, and by solving the constrained optimization problem, we obtain the most probable $s(\mathrm{BE})$ with two fully separated features, as shown in Fig.~\ref{suppfig:sBEresults} . 
This convolution model completely relies on the accurate calibration of the kinetic energy axis and photon energy axis, 
which has been done with our main approach with the slope of the photoelectron feature in the resulting $I_0(E,\omega)$.  
As discussed in the main text, the comparative advantages of these related approaches may be case dependent, and the precise choice of method depends critically on the properties and purpose of the measurements. %

\bibliographystyle{apsrev4-1}
\bibliography{main.bib, referencesJPC.bib}

%% file: defs.tex
\newcommand{\cov}{\mathrm{Cov}}

\renewcommand{\d}{\mathrm{d}}